\documentclass[floats,floatfix,showpacs,amssymb,amsmath,prd,twocolumn,superscriptaddress,nofootinbib, aps]{revtex4-2}

\usepackage[utf8]{inputenc}
\usepackage[T1]{fontenc}
\usepackage{hyperref}
\usepackage{bm}
\usepackage{dcolumn}
\usepackage{aas_macros}
\usepackage{amsmath, amssymb}
\usepackage{graphicx}
\usepackage{float}
\usepackage[normalem]{ulem}
\usepackage{soul}
\setstcolor{red}
\hypersetup{colorlinks,bookmarksnumbered,
citecolor=cyan, urlcolor=magenta}

\providecommand{\adsurl}[1]{\href{#1}{ADS}}
\usepackage{natbib}
\usepackage[usenames, dvipsnames]{xcolor}

\begin{document}
\preprint{APS/123-QED}

\title{Constraining the Stochastic Gravitational Wave Background with Photometric Surveys}

\author{Yijun Wang}
\email{yijunw@caltech.edu}
\affiliation{California Institute of Technology, Pasadena, CA 91125, USA}
\author{Kris Pardo}
\affiliation{Jet Propulsion Laboratory, California Institute of Technology, Pasadena, CA 91109, USA}
\author{Tzu-Ching Chang}
\affiliation{Jet Propulsion Laboratory, California Institute of Technology, Pasadena, CA 91109, USA}
\affiliation{California Institute of Technology, Pasadena, CA 91125, USA}
\author{Olivier Dor\'e}
\affiliation{Jet Propulsion Laboratory, California Institute of Technology, Pasadena, CA 91109, USA}
\affiliation{California Institute of Technology, Pasadena, CA 91125, USA}

\date{\today}
\begin{abstract}

The detection of the Stochastic Gravitational Wave Background (SGWB) is essential for understanding black hole populations, especially for supermassive black hole binaries. The recent promising results from various Pulsar Timing Array (PTA) collaborations allude to an imminent detection. In this paper, we investigate the relative astrometric gravitational wave detection method, which can contribute to SGWB studies in the microhertz range. We consider the Roman Space Telescope and \textit{Gaia} as candidates and quantitatively discuss the survey sensitivity in both the frequency and spatial domains. We emphasize the importance of survey specific constraints on performance estimates by considering mean field of view (FoV) signal subtraction and angular power spectrum binning. We conclude that if the SGWB is at a similar level as in PTA estimates, both Roman and \textit{Gaia} have the potential to detect this frequency-domain power excess. However, both Roman and \textit{Gaia} are subject to FoV limitations, and are unlikely to be sensitive to the spatial pattern of the SGWB. 

\end{abstract}
\maketitle

\section{introduction}
\label{sec:intro}
The successful detections of compact binary coalescence (CBC) events by the LIGO and Virgo collaborations have opened the possibility of gravitational wave (GW) astronomy \citep[see, e.g.,][]{GW150914,GW170817,GW190412,GWTC2}. Sensitive to GWs in the $10 \sim 1000 $ Hz range, the aLIGO network and other proposed ground-based next-generation GW detectors are at the prime frequency range to detect individual mergers between stellar-mass compact objects \citep[see e.g.,][]{Voyager2020,CE2019,ET2011}. To observe GW signals from higher-mass systems requires detector coverage in lower frequency ranges. In the decihertz to millihertz range, a series of space-based interferometer-type GW detectors, such as LISA \cite{LISAL3}, DeciGO \cite{DECIGO2019} and TianQin \cite{TianQin2016}, will target signals from binaries with masses between $100\sim 10^7~M_{\odot}$ \citep[see, e.g.,][]{LISAL3,TianQinScience,decigoscience}. The nanohertz range is covered by the Pulsar Timing Array (PTA) method, championed by several collaborations including the North American Nanohertz Observatory for Gravitational Waves (NANOGrav) \cite{nanograv_concept}, the European Pulsar Timing Array (EPTA) \cite{epta_concept}, and the Parkes Pulsar Timing Array (PPTA) \cite{ppta_concept}. 

In addition to individual CBCs, an important detection candidate is the stochastic gravitational wave background (SGWB). A prominent SGWB source is the superposition of GWs from supermassive black hole binaries (SMBHBs) throughout cosmic history. While individual GWs can be too weak to be detectable, the combination can be sufficiently loud; indeed, it is predicted that PTAs will detect the SGWB prior to individual GW detections \citep[see, e.g.,][]{Rosado}. Other contributing sources to the SGWB have also been proposed, such as from cosmic strings and phase transitions in the early universe \cite{Chang_cosmicstring,Siemens}.

Recently, the NANOGrav 12.5-yr data analysis detected a common red noise among the observed pulsars with an amplitude of $1.92\times 10^{-15}$ at the reference frequency $1~{\rm yr}^{-1}$ ($3.2\times10^{-8}$ Hz) with high confidence \cite{nanograv12yr}. Shortly afterwards, analysis of the 24-yr EPTA dataset revealed a common red noise with an amplitude of $2.95\times 10^{-15}$ at the same reference frequency, in rough agreement with the NANOGrav result \cite{EPTA_det}. Similar detection results have since been published by PPTA \cite{PPTA} and IPTA \cite{IPTADR2}. While these studies show insufficient evidence of the signal angular correlation, famously known as the Hellings-Downs curve \cite{Hellings}, to positively identify the signal as the SGWB, they suggest the prospect of an imminent detection with further data collection. 

While PTAs are exceptionally suited for detection in the nanohertz range, there are currently no ongoing or planned observatories to cover the SGWB in the microhertz band. In this range, we expect that the SGWB will be produced by lighter binaries ($10^5\sim 10^9~M_\odot$) than those seen by PTAs \citep[see, e.g.,][]{Sesana_mu}. Observing the SGWB in this regime would complement and cross-check the PTA observations, as the SGWB at different frequencies should eventually be consistent with the same population model. Recently, several potential detection methods have been proposed that target this uncovered frequency band gap. For example, GW could be detected via its modifications to asteroid accelerations \cite{Fedderke}. It also modulates the phase of continuous GWs from galactic sources \cite{Rosell}, from which the SGWB can be inferred. Spaced-based interferometer-type detectors are also proposed \cite{Sesana_mu}. 

In this paper, we focus on the detection method using relative astrometric measurements \cite{Moore2017,Klioner2018,Wang}. Analogous to the periodic pulse arrival time delay in PTA, the sky position of measured objects oscillates with a passing GW, from which we infer the source properties \cite{Book2011,Pyne1996}. In particular, the astrometry method has the advantage of being flexible in its sensitive frequency range with no additional instrument cost to the photometric surveys \cite{Wang}. By specifically referring to photometric surveys, we wish to emphasize that not only surveys dedicated to astrometry can achieve this purpose; indeed, a promising candidate, the Galactic Bulge Time Domain (GBTD) survey of the Nancy Grace Roman Space Telescope, has the primary objective to observe microlensing signatures in the galactic bulge \cite{GBTDwebsite}. 

Several authors have investigated the potential of this method, both from a theoretical perspective \cite{Book2011,Qin,MihaylovNonC,MihaylovNonE,Golat} and in close connection with specific surveys \cite{Moore2017,Klioner2018,Wang}. In Ref.~\cite{Wang}, we investigate the detection prospect of individual monochromatic GWs in the case of the Nancy Grace Roman Space Telescope (Roman) \cite{Roman} and \textit{Gaia} \cite{Gaia2016}, both offering high-precision and high-cadence astrometric measurements. 

In this work, we elaborate on the effect of survey features on the sensitivity to SGWB. Explicitly, we compare surveys depending on whether they are sensitive to a largely uniform displacement of all stars within their field of view (FoV). We clarify that, although the detection method itself requires relative astrometry only, the design for absolute astrometry telescopes contributes to higher sensitivity. We also discuss detection implications due to the size of the FoV; we apply an angular power spectrum binning and compute the recoverable signal power for given FoV size.

This paper is organized as follows. In Sec.~\ref{sec:theory}, we review the fundamentals of GW-induced astrometric deflections. We then develop the expected power spectra in both the frequency domain and the spatial domain. In Sec.~\ref{sec:analysis}, we examine the respective survey sensitivities under various survey features. We conclude in Sec.~\ref{sec:conclusion}.

The code used for this analysis and the figures is available at \href{https://github.com/kpardo/estoiles-public}{https://github.com/kpardo/estoiles-public}.

\section{Theory}
\label{sec:theory}
In this section, we review the basics of GW-induced astrometric deflections. We then discuss two different ways to analyze the deflections produced by the SGWB: the frequency power spectrum and the angular power spectrum. We then show the corresponding instrument noise power spectrum. We emphasize, however, that these signal and noise prescriptions alone are insufficient to gauge the GW detection power of a photometric survey; as we will show in Sec.~\ref{sec:analysis}, a more accurate description is contingent on additional survey features. 

\subsection{GW Detection with Astrometry}

Upon the passing of a GW, the propagation path of photons are perturbed such that the measured star positions are deflected from their ``true'' positions. The deflection vector $d\vec{n}$ of the position of a star located at $\vec{n}$ on the celestial sphere is given as \cite{Book2011,Pyne1996} 

\begin{equation}
\label{dneqn}
    dn^i = \frac{n^i+p^i}{2(1+p_jn^j)}h_{jk}n^jn^k - \frac{1}{2}h^{ij}n_j\;,
\end{equation}
where $\vec{p}$ is the propagation direction of the GW, $h_{ij}$ is the GW strain tensor evaluated at the observer, and the Latin index ranges over the three spatial dimensions. We adopt the Einstein notation, in which repeated upper and lower indices imply summation. In Eqn.~\eqref{dneqn}, we adopt the distant source limit, where the distance to the GW source and to the observed stars are much larger than the GW wavelength. Consequently, the GW perturbations at different star locations are uncorrelated, and can be treated as a source of random noise \cite{Book2011,Pyne1996}. In this work, we are interested in GWs with frequencies ranging from $10^{-8}\sim10^{-5}$ Hz where this distant source limit is almost always valid. Given the observable $d\vec{n}$, it is then possible to infer the source property, i.e., $h_{ij}$.

The astrometry GW detection method has several merits. Firstly, it is highly versatile in its sensitive frequency range, given as
\begin{equation}
    f_{\rm min}\sim \frac{1}{T_{\rm obs}},~ f_{\rm max}\sim \frac{1}{\Delta t}\;,
\label{eqn:f_range}
\end{equation}
where $T_{\rm obs}$ is the survey lifetime, and $\Delta t$ is the observational cadence. Therefore, surveys with the right cadence can potentially bridge the microhertz GW frequency spectrum. 

In addition, the detection sensitivity is boosted by the number of observed stars and the number of exposures for each star. Given suitable surveys, these factors are typically quite large; for example, in the Galactic Bulge Time Domain Survey (GBTD) of Roman, each of the $\sim 10^8$ observed targets has up to $4.1\times 10^4$ exposures \cite{Sanderson2019}. 

Furthermore, this detection method requires astrometric measurements only, without additional equipment and observing time; in this way, the GW scientific output is serendipitous given existing surveys. Finally, we note that Eqn.~\eqref{dneqn} suggests that only relative astrometric measurements (i.e., away from \textit{some} fixed reference location) are needed; it is not necessary that we know the absolute astrometric coordinates. We discuss this point in detail in Sec.~\ref{sec:analysis}. In the following sections, unless explicitly stated, the astrometric measurements are to be interpreted as relative ones. 

\subsection{Frequency Power Spectrum}

One way to analyze the astrometric data is to compute the signal magnitude correlation across time, which can be represented by its Fourier transform, i.e., the SGWB frequency-domain power spectrum \citep[see, e.g.,][]{Phinney2001,Allen_SGWB}.

The SGWB follows the assumption that the time-domain strain amplitude follows a stationary zero-mean Gaussian distribution, and its Fourier transform is a sum of Gaussian modes with frequency-dependent variance, i.e.,
\begin{equation}
\label{eqn:hvar}
\begin{split}
    \langle h(t)h(t')\rangle &= \delta^D(t-t')\sigma_t^2\\
    \langle \tilde{h}(f)\tilde{h}^*(f')\rangle &= \delta^D(f-f')S_h(f)\;,
\end{split}
\end{equation}
where $\delta^D$ is the Dirac delta and $\sigma_t$ is the constant standard deviation of the strain. The double-sided signal power spectrum, $S_h(f)$, can be expressed as the GW energy density per logarithmic frequency interval, $\Omega_{\rm gw}(f)$, or the characteristic strain amplitude, $h_c(f)$, with the relationship given as \cite{Phinney2001}:

\begin{equation}
\label{eqn:EhcSh}
\begin{split}
    \rho_cc^2\Omega_{\rm gw}(f) &= \frac{\pi c^2}{4G}f^2h_c^2(f) = 2\frac{\pi c^2}{4G}f^3S_h(f)\;,
\end{split}
\end{equation}
and the critical density today is given by 
\begin{equation}
    \rho_c = \frac{3H_0^2}{8\pi G}\;,
\end{equation}
where $H_0$ is the Hubble constant today. 

The above formulae are independent from the specific form of the characteristic strain amplitude (or equivalent quantities). A generic phenomenological model for a source-agnostic SGWB is a power law,
\begin{equation}
\label{hcagnostic}
    h_c(f) \sim A\left(\frac{f}{f_{\rm ref}}\right)^\alpha\;,
\end{equation}
completely specified by the spectral slope $\alpha$ and the spectrum amplitude, $A$, at a reference frequency $f_{\rm ref}$. The slope, in particular, is determined by the nature of the SGWB source. For example, the SGWB from inspiralling black hole binaries has a spectral slope of $-2/3$ \cite{Phinney2001,Cutler_SPA}; for primordial background (e.g. inflation) and cosmic strings, the spectral slopes are $-1$ and $-7/6$, respectively \cite{Grishchuk, Damour_string,EPTA_cosmicstring}. 

For a survey taking $N_m$ exposures of $N_s$ stars, the \textit{effective} single-exposure single-star noise variance is given by an ensemble average as
\begin{equation}
\label{noisevar}
    \langle n_{IJ}n_{I'J'}\rangle =  \frac{\sigma_{ss}^2\delta^K_{II',JJ'}}{N_sN_m}\;,
\end{equation}
where $I,J$ are indices for the exposure and the star and $\sigma_{ss}$ is the single-star single exposure astrometric noise standard deviation. The symbol $\delta_K$ is the Kronecker delta. We assume the measurement noise is both spatially and temporally uncorrelated. In analogy to the GW power spectrum, we define the noise power spectrum as
\begin{equation}
    \langle n_{IJ}n_{I'J'}\rangle\equiv\delta^D(t_I-t_{I'})P_n^{}\delta^K_{JJ'}\;.
\end{equation}
For a finite observation time, the discrete version of the Dirac delta is given by 
\begin{equation}
    \delta^D(t_I-t_{I'}) \rightarrow \frac{1}{\Delta t}\delta^K_{II'}\;,
\end{equation}
where $\Delta t$ is the (constant) time interval between exposures. It follows that 
\begin{equation}
    P_n = \Delta t\frac{\sigma_{ss}^2}{N_sN_m}\;.
\label{Pnfreq}
\end{equation}

Since the deflection signal $|d\vec{n}|$ and the GW amplitude $|h(t)|$ differ only by an $\mathcal{O}(1)$ geometric factor, we make the approximation that $|d\vec{n}|\sim |h(t)|$, and Eqn.~\eqref{Pnfreq} is approximately the strain noise power spectrum. The corresponding characteristic amplitude is given by 
\begin{equation}
    h_n(f) = \sqrt{2fP_n} = \sigma_{ss}\sqrt{\frac{2f\Delta t}{N_sN_m}}\;,
\label{hn}
\end{equation}
where we insert a factor of 2 to convert from a double-sided power spectrum to single-sided. For a survey with fixed cadence, $h_n(f)$ scales as $1/\sqrt{T_{\rm obs}}$, consistent with Ref.~\cite{Thrane}. 

\subsection{Angular Power Spectrum}

Aside from looking at time stream data, we may take any exposure and concentrate on the angular correlations between deflection signals at different sky locations at a given time. This method has been very broadly applied; for example, in the PTA search for SGWB, this spatial correlation is expressed in the form of the Hellings and Downs curve \cite{Hellings}. A similar strategy has been applied to studies of the Cosmic Microwave Background (CMB), galaxy distributions and dark matter distributions \citep[see, e.g.,][]{Hu_wl_cmb,Nicola_subaru,Legrand_gal,Okamoto,Zaldarriaga}. We note that the most commonly analyzed signals are spin-0 scalar signals, (e.g., CMB temperature map, galaxy counts) and spin-2 tensor signals, (e.g., CMB polarization, weak lensing distortion map). The GW-induced astrometric deflection is a spin-1 vector signal, and will be decomposed with vector spherical harmonics as  \cite{Qin,Golat,MihaylovNonE,Okamoto,Hu_wl_cmb}

\begin{equation}
\label{dnEB}
    (d n)_a = \sum_{\ell m}\left[E_{\ell m}Y^E_{(\ell m),a}(\vec{n}) + B_{\ell m}Y^B_{(\ell m),a}(\vec{n})\right]\;,
\end{equation}
where $a$ indicates two orthogonal unit vectors tangential to the celestial sphere. The E,B basis is chosen such that the coefficients $E_{\ell m}(B_{\ell m})$ transform as scalars (pseudo-scalars) under a local rotation \cite{ZaldarriagaEB}. 

We assume that the complex coefficients $E_{\ell m},B_{\ell m}$ are drawn from zero-mean normal distributions, with the variance given by 
\begin{equation}
    \begin{split}
        \langle E_{\ell m}E^*_{\ell'm'} \rangle &= C^E_\ell \delta_{\ell \ell',mm'}\\
        \langle B_{\ell m}B^*_{\ell'm'} \rangle &= C^B_\ell \delta_{\ell\ell',mm'}\;.
    \end{split}
\end{equation}

Just as the characteristic strain amplitude, $h_c(f)$, indicates the SGWB source, $C^{E,B}_{\ell}$ depends on the nature of GW radiation itself, such as its propagation speed and polarization content \cite{Qin,MihaylovNonC,MihaylovNonE}. In this work, we assume the GW travels at the speed of light and contains only tensor modes. The angular power spectrum is then given as \cite{Qin}

\begin{equation}
\label{eqn:Cell}
\begin{split}
    C^E_\ell = C^B_\ell &= \frac{12H_0^2N_\ell^{-2}}{\pi \ell (\ell+1)}\int df~\frac{\Omega_{\rm gw}(f)}{f^3}|W(f)|^2\\
    &= \frac{8\pi N_\ell^{-2}}{\ell(\ell+1)}\int d\ln f~ h_c^2(f)\\
    N_\ell&\equiv\sqrt{\frac{(\ell+2)!}{2(\ell-2)!}} \;.
\end{split}
\end{equation}
The window function $W(f)$ accounts for the phase difference between two exposures. This factor arises, as the model in Ref.~\cite{Qin} assumes two exposures only, and the ``deflection'' must be calculated using one of the two exposures as the baseline. In our model, we assume that a true baseline is established by averaging the measurements over the entire observational period. In this way, $W(f)$ is not necessary. We observe that the angular power spectrum is sharply peaked at small $\ell$ and rapidly drops off as $\ell^{-6}$ at large $\ell$.

From an observational perspective, the angular power spectra $C^{E,B}_{\ell}(t_I)$ at each time slice $t_I$ can be extracted via the inverse of Eqn.~\eqref{dnEB}. Since this angular power spectrum is stationary, the final estimated angular power spectrum can be averaged over the exposures,
\begin{equation}
    C^{E,B}_{\ell} = \frac{1}{N_m}\sum_I C^{E,B}_{\ell}(t_{I})\;.
\end{equation}
This formula can also be understood from the definition of $C^{E,B}_\ell$ as the ensemble average of $E,B_{\ell m}$; for a single exposure the average runs over the $m$ modes, and for multiple exposures it also runs over the (independent) realizations of the angular power spectrum.

The spatial measurement noise is modeled as vectors on the two-sphere with random orientations, with magnitude drawn from a normal distribution as in Eqn.~\eqref{noisevar}. Applying a harmonic transform, we obtain the noise power spectrum as
\begin{equation}
    C^{E,B}_n = \frac{2\pi\sigma_{ss}^2}{N_mN_s}\;,
\end{equation}
where the factor of $2\pi$ reflects the angular normalization and that the power is split evenly between the $E,B$ modes \cite{Qin}.

\section{Analysis}
\label{sec:analysis}
While Sec.~\ref{sec:theory} provides the theoretical signal power spectra, they are not necessarily representative of what can be recovered from observational data. In this section, we clarify necessary modifications to the power spectra such that they are applicable for specific types of photometric surveys. Firstly, we describe our reference surveys, the Roman GBTD survey and the \textit{Gaia} astrometric survey. We then discuss the role of absolute and relative astrometry in GW detection. We discuss in depth the effect of subtracting the FoV mean signal and the resulting survey performance. Lastly we introduce angular power binning and explain the implication of limited FoV on GW detection. 

\subsection{Survey Summary}

The Nancy Grace Roman Space Telescope\footnote{https://roman.gsfc.nasa.gov/} is NASA's next flagship observatory after the James Webb Space Telescope. Among other science goals, it aims to probe the evolution of dark energy and large-scale structure by observing billions of galaxies and thousands of supernovae \cite{Agrawal}. In terms of probing GWs, the GBTD survey is particularly relevant, where it visits a $\sim 1.97~ {\rm deg}^2$ patch of sky towards the galactic center. This pointing direction implies large stellar density, and the near infrared sensitive wavelength also leads to less extinction; thanks to these factors, Roman is expected to observe $10^8$ stars ($W145_{AB}<23$) \cite{Gaudi2019} with the GBTD survey, with a single-exposure single-star astrometric uncertainty of 1.1 mas (estimated for $H_{\rm AB}=21.6$ stars) \cite{Sanderson2019}. The survey comprises of 6 observing seasons, each 72 days long. During each season, each star is visited every 15 minutes, giving a total of 41,000 exposures per star. 

Another promising mission as a potential GW probe is the all-sky astrometric survey with \textit{Gaia}\footnote{https://sci.esa.int/web/gaia} \cite{Gaia2016}. \textit{Gaia} observes on the order of $10^9$ stars \cite{GaiaDR2_2018}, with a single-exposure single-star astrometric uncertainty of 0.7 mas for $G\sim20$ stars in \textit{Gaia} Data Release 2 \cite{Luri2018}. In the recent \textit{Gaia} Early Data Release 3, the typical uncertainty for six-parameter astrometry (position) is 0.4 mas at $G=20$ \cite{GaiaEDR3}. \textit{Gaia} Data Release 4, which will be based on data during the entire nominal mission lifetime and part of the extended mission, expects a parallax uncertainty of 0.46 mas at $G=20$ \cite{GaiaFutureDR}. \textit{Gaia} also expects to achieve a ten-year total lifetime in the extended mission \cite{Brown}; the ultimate Data Release 5 expects to have a parallax uncertainty of 0.33 mas for objects of the same magnitude \cite{GaiaFutureDR} (the astrometric position uncertainty generally has a small difference in value with parallax uncertainty \citep[see, e.g.,][]{GaiaEDR3}). Since the difference is within an order of magnitude, we do not update the estimates accordingly, and the general conclusion remains the same. On average, each star is visited 70 times throughout the 5-year observing time \cite{Gaia2016}. For simplicity, we assume these exposures to be evenly spaced. 

\subsection{Absolute/Relative Astrometry and Mean Subtraction}

As is evident from Eqn.~\eqref{eqn:hvar}, the SGWB inference relies solely on the deflection vectors and not on the true position vector; indeed, if we observe cycles of a single continuous wave, or a significant period of the stochastic process, the true position can be immediately computed. Therefore, it is not required to have absolute astrometric measurements (as \textit{Gaia} provides), consisting of the absolute star coordinates in, e.g., the extragalactic International Celestial Reference System \cite{GaiaAstrometry}.

What is required, on the other hand, is that the entire deflection vectors should be recovered from the time stream data. This is particularly challenging; since the GW deflection pattern is a large scale signal (see Eqn.~\eqref{eqn:Cell}), deflections within a small FoV appear almost uniform. We shall refer to this almost uniform motion as the FoV mean signal.

For telescopes with a single viewing direction, the mean signal is recoverable under two possibilities. Evidently, a telescope that is in free fall during data collection qualifies, since an inertial frame cannot ``absorb'' periodic motion. For a ``point-and-stare'' type telescope like Roman, its reaction wheels are constantly engaged through the fine guidance system for pointing self-calibration \cite{Sanderson2019}. As the mean signal mimics instrument noises such as pointing error and jitter, it may then be corrected \textit{in situ} or get fitted out during the astrometric solution process. In this case, it is in principle possible to reconstruct the mean signal (or part of it) if the pointing system actions are recorded. 

In the worst case scenario, we ignore the mean signal and use only the differential deflections across each exposures, which we refer to as the mean-subtracted case. In Ref.~\cite{Wang}, we considered this case for individual continuous wave detection and showed that the differential deflections are roughly two orders of magnitude smaller than the mean signal. The sensitivity loss can be approximated by scaling down the signal characteristic strain amplitude proportionally.

For telescopes with two almost orthogonal viewing directions, such as \textit{Gaia}, the GW signature is typically towards two distinctive directions in the two viewing directions. In this case, the signal is less likely to be absorbed as pointing error. Although the \textit{Gaia} viewing directions were not specifically designed for this purpose, it incidentally satisfies the favoring conditions. 

In the following, we consider the signal-to-noise ratio (SNR) defined as \citep[see, e.g.,][]{Moore2017}
\begin{equation}
    \rho^2 = \int d\ln f~ \left[\frac{\lambda h_c(f)}{h_n(f)}\right]^2\;,
\end{equation}
where $h_c$ is the SGWB characteristic strain amplitude and $h_n$ is the characteristic noise amplitude in Eqn.~\eqref{hn}, and the sensitivity threshold is fixed to have $\rho=1$. The scaling factor $\lambda$ accounts for the signal loss due to mean subtraction. For \textit{Gaia}, we assume $\lambda=1$, i.e. lossless. As the Roman telescope and system design is still in planning, we do not quote a specific value; instead, we show the range $\lambda = 0.01\sim 1$, which corresponds to the mean-subtracted case and the full signal case, respectively \cite{Wang}. 

\begin{figure}
    \includegraphics[width=\columnwidth]{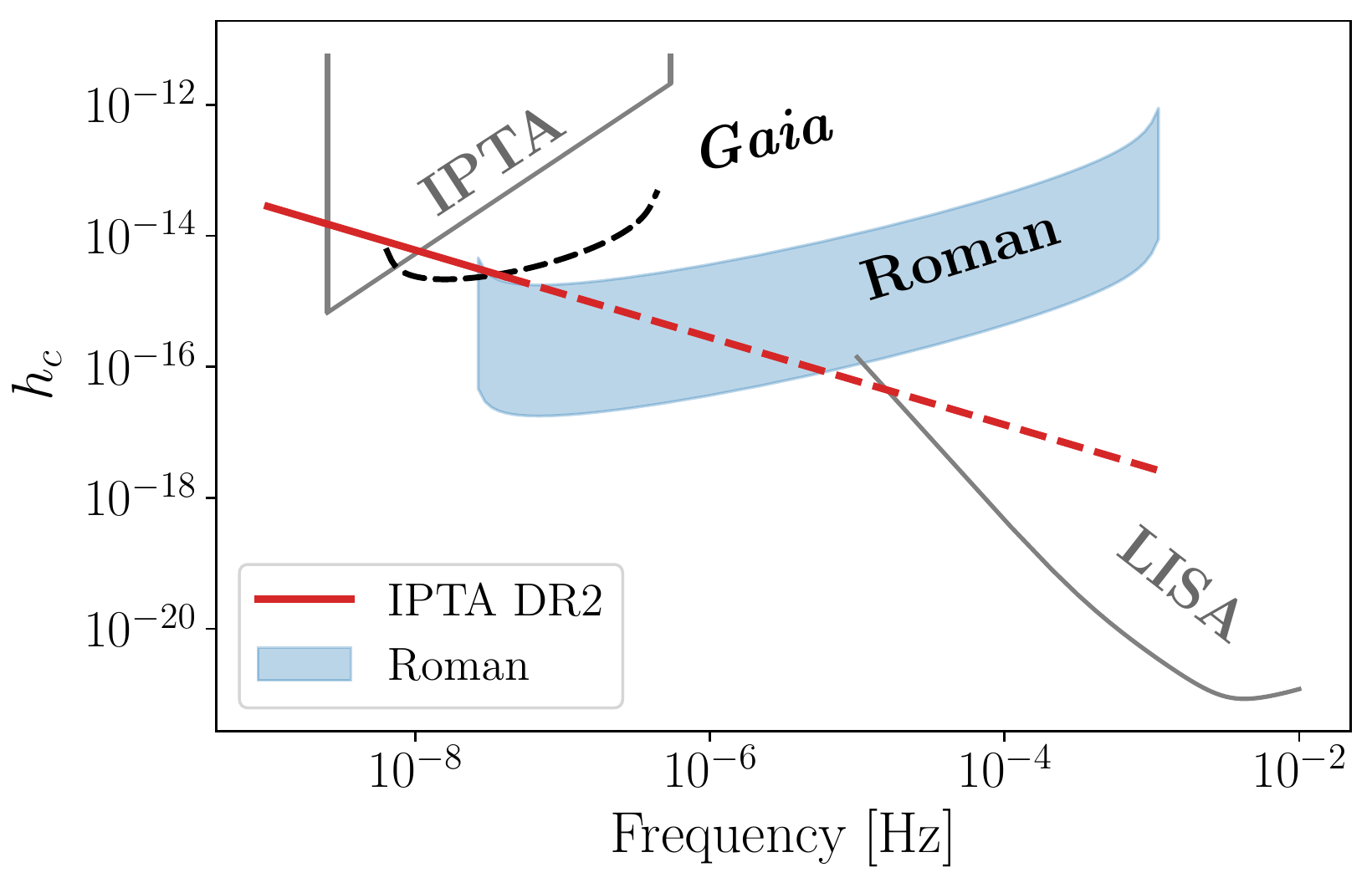}
    \caption{Survey sensitivity for a source-agnostic SGWB search. The survey sensitivity of IPTA and LISA are shown in gray curves. The solid black line shows sensitivity of \textit{Gaia}. The range of sensitivity of Roman GBTD survey is shown with the blue solid block, with $\lambda=0.01\sim 1$. The solid red line shows the best amplitude estimate from IPTA DR2, with the same frequency range as in Fig. 1 of Ref.~\cite{IPTADR2}. The dashed red line shows the extrapolated SGWB level in the Roman frequency range.} 
    \label{fig:hcsens}
\end{figure}

We show the sensitivity threshold in Fig.~\ref{fig:hcsens}, where we do not restrict to the SMBHB SGWB, but rather assume a source-agnostic search \cite{Thrane}. For reference, we also plot the sensitivity curves of IPTA \cite{Taylor2016,Verbiest2016} and LISA \cite{Robson2019} in gray using phenomenological models. The solid red line shows the best estimate of the common process measured from IPTA DR2 (over $10^{-9}- 4\times 10^{-8}~\rm Hz$, consistent with Fig. 1 of Ref.~\cite{IPTADR2}),
\begin{equation}
    h_{c,\rm IPTA} \sim 2.8\times 10^{-15}\left(\frac{f}{1~{\rm yr}^{-1}}\right)^{-2/3}\;.
\end{equation}
We note that the population details of potential SMBHBs emitting GWs from $10^{-7}$ Hz to $10^{-4}$ Hz are highly uncertain and poorly constrained by observation, if at all. As a heuristic example, we simply extrapolate the SGWB from the nanohertz range (red dashed line). While this treatment is valid until some of the more massive GW sources go near coalescence, it suffices in this work as an example to possible SGWB signals across the wide frequency ranges.

We reiterate that the frequency band difference between Roman and \textit{Gaia} results from the observational cadence (see Eqn.~\eqref{eqn:f_range}). The sensitivity, in addition to the cadence, is affected by the astrometric accuracy and the number of observed stars. Specifically, we observe that within shared frequency ranges, Roman offers better sensitivity than \textit{Gaia}, even in the mean-subtracted scenario. This is primarily due to its high cadence and larger span in its sensitive frequency. We note that if the SMBHB SGWB is indeed at a similar level as the current best estimate from IPTA, both \textit{Gaia} and Roman can detect its power excess. 

Given a fixed redshift distribution, the SGWB amplitude directly implies the local SMBHB remnant density, $\Phi_{\rm BHB,0}$, defined as the number of SMBHB remnant per comoving volume in the local universe. Assuming the same SMBHB population model as in Ref.~\cite{IPTADR2,CaseyClyde2022}, we can also express the Roman and \textit{Gaia} sensitivity in terms of an upper limit on the local SMBHB number density (in the case of null detection), since $A\sim \sqrt{\Phi_{\rm BHB,0}}$ in Eqn.~\eqref{hcagnostic}. For Roman, the upper limit is $1.0\times 10^{-5}\sim1.0\times 10^{-9}~{\rm Mpc}^{-3}$ for $\lambda=0.01\sim 1$. For \textit{Gaia}, the upper limit is $2.3\times 10^{-6}~{\rm Mpc}^{-3}$.

\subsection{Angular Power Binning}

In this section, we discuss the effect of telescope FoV size on the angular power spectrum sensitivity. Firstly, we note that the currently planned telescopes with sufficient astrometric precision for GW measurement are all space-based telescopes with FoV less than 1 deg$^2$; e.g., Roman (0.26 deg$^2$) \cite{Sanderson2019}, \textit{Gaia} (0.5 deg$^2$) \cite{GaiaFoV}, James Webb Space Telescope ($\sim 8~{\rm arcmin}^2$)\footnote{https://www.jwst.nasa.gov/},\footnote{https://svs.gsfc.nasa.gov/13583}, Hubble Space Telescope (WFC3 NIR, $\sim 4~{\rm arcmin}^2$)\footnote{https://hubblesite.org/}. Therefore, it is not likely that we can capture GW deflections with significant spatial variation within one exposure. We further discuss in turn the ``point-and-stare'' type surveys (e.g., Roman) and all-sky scanning surveys (e.g., \textit{Gaia}).

In the case of Roman GBTD survey, which only visits fields close to the galactic center, the measurements are insensitive to angular powers with a scale larger than the FoV size\footnote{In principle, measurements from the several fields Roman visits may be combined on particular time scales (e.g., much shorter than the typical SGWB period of interest). Such reconstruction would require careful modeling of the telescope motion during field-switching; for simplicity, we restrict to scales smaller than the FoV.}. Moreover, different $(\ell,m)$ modes of the spherical harmonic decomposition coefficients can be highly coupled. While the exact coupling depends on the FoV geometry, modes with larger difference in value than $\pi/\theta_{\rm fov}$
have much smaller coupling, where $\theta_{\rm fov}$ is the (angular) sidelength of the FoV. 

Although \textit{Gaia} offers complete sky coverage after each full-sky scan, it is unclear if it will not suffer the same angular power loss. In the case of the SGWB, the spatial deflection patterns at different times are, by assumption, independent realizations of $C_\ell^{E,B}$. The image of the FoV are also independent, as a consequence. In this way, measurements from different exposures cannot be consistently combined to produce a full-sky map, and \textit{Gaia} suffers the same large-scale power mean-signal loss as Roman.

In the case of individual continuous GW signals, the signal coherence allows, in principle, the construction of a temporal-spatial template, which depends on the GW source property and the attitude history of the telescope \cite{GaiaAstrometry}. A thorough investigation of such a possibility is beyond the scope of this work. In the following analysis, we shall assume that the large-scale powers cannot be recovered from \textit{Gaia} measurements. 

To account for the large-scale power loss and mode mixing, we bin the theoretical angular power spectrum and impose a minimum $\ell$. The maximum $\ell$ roughly corresponds to pixel scale, which is much larger than any $\ell$-modes contributing significantly to the SGWB, i.e., $\ell\sim 2$. The binned, predicted angular power spectrum can be obtained via the exact mode-coupling matrix of the FoV (i.e., the window) \cite{Alonso,Hivon,Wandelt}; however, to simplify the calculation and to keep the estimate applicable to generic FoV shapes, we approximate the binning process by directly averaging the theoretical $C^{E,B}_\ell$ within each bin \citep[see, e.g.,][]{Hivon}. Given sufficiently wide bins, different bins have negligible coupling.

We consider a square FoV, where the bins are defined as 
\begin{equation}
\label{eqn:lbin}
\begin{split}
    &C^{E,B}_{q} = \frac{1}{\Delta q}\sum_{\ell_i\in \{\ell_q\}}C^{E,B}_{\ell_i},~ \Delta q = \left[\frac{\pi}{\theta_{\rm fov}}\right],\\
   & \{\ell_q\}=\{\ell_i,\left[q-\frac{1}{2}\right]\Delta q\leq i<\left[q+\frac{1}{2}\right]\Delta q\}\;,
\end{split}
\end{equation}
where $\left[\cdot\right]$ denotes taking the nearest integer and $\theta_{\rm fov}$ is the angular sidelength of the square FoV.

Since the survey is not sensitive to modes larger than the FoV, we discard the first bin when calculating the SNR. With these two steps, the log likelihood is summed over the binned modes as \citep[see, e.g.,][]{Hu_wl_cmb}
\begin{equation}
\begin{split}
    \rho_{E,B}^2 &= \sum_q \left(\frac{C^{E,B}_q}{\Delta C^{E,B}_q}\right)^2\\
    \rho^2 &= \rho_E^2+\rho_B^2\;.
\end{split}
\end{equation}
In the weak signal limit, the diagonal elements of the covariance matrix, $1/\Delta C^{E,B}_q$, are dominated by the noise, instead of cosmic variance. The inverse of the covariance for each bin is given by \cite{Dodelson}
\begin{equation}
    \Delta C^{E,B}_q =  \frac{1}{\Delta_q}\sqrt{\frac{2}{(2\ell_q+1)f_{\rm sky}}}C^{E,B}_n\;,
\end{equation}
where $\ell_q$ is the mode in the center of each bin. The factor $f_{\rm sky}$ if the fraction of sky covered by the FoV, and it accounts for the loss of mode power due to partial sky coverage. 

It is clear that the SNR loss due to limited FoV is also sensitive to the signal spectral slope; signal concentrated at large scales suffers a higher loss of SNR compared with signals that have a flatter spectral profile. As Eqn.~\eqref{eqn:Cell} shows, the angular power spectrum of SGWB decays roughly as $\ell^{-6}$, which strongly penalizes the detecting power of limited-FoV surveys. 

\begin{figure}
    \includegraphics[width=\columnwidth]{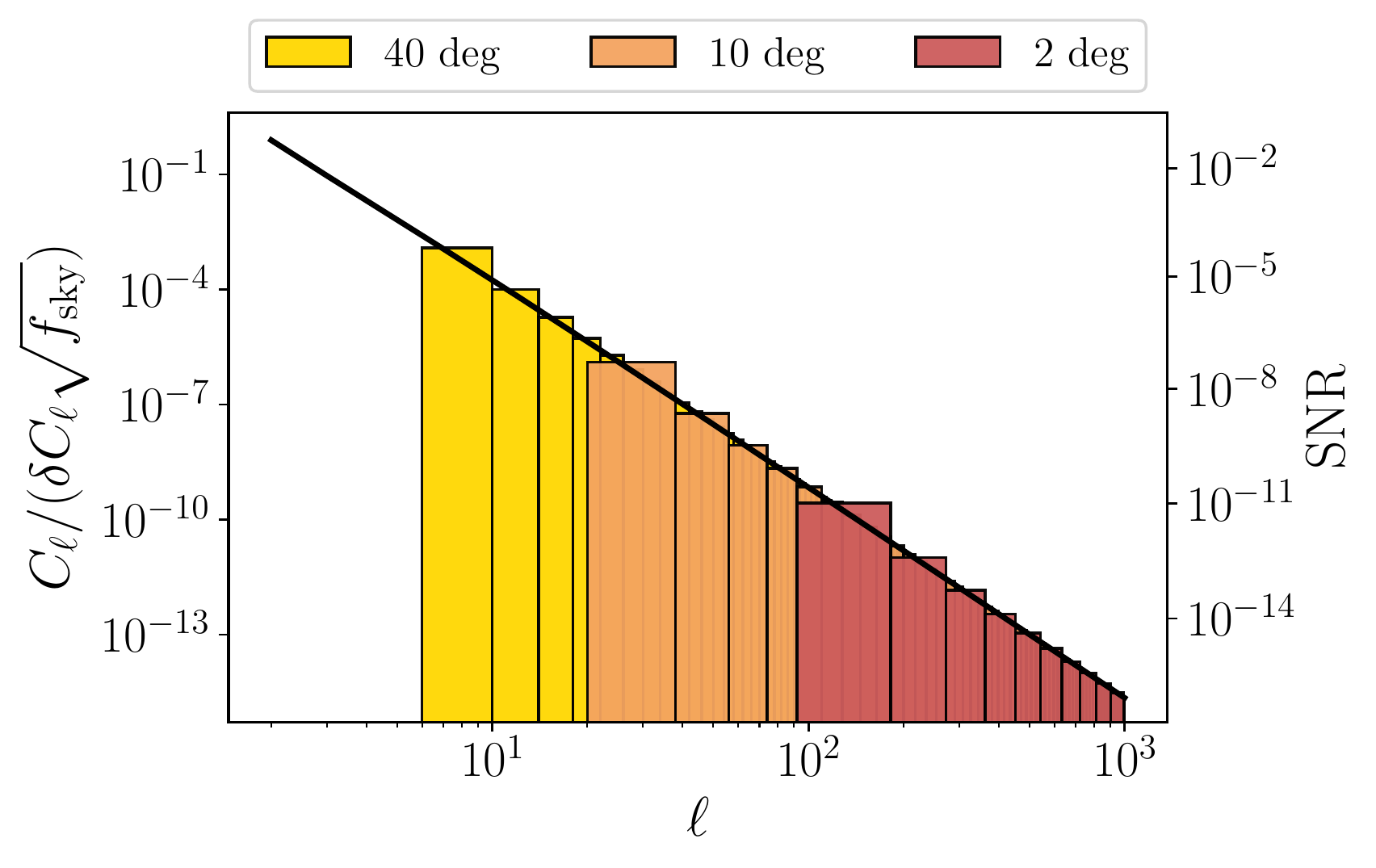}
    \caption{The fractional SNR for various FoV size compared with the full sky case. The bar plot corresponds to the left axis and shows the SNR contribution from each bin, normalized by $f_{\rm sky}$. The black solid line shows the full-sky scenario. Using the right axis, the same line shows the total SNR with various threshold $\ell=\pi/\theta_{\rm fov}$ as a fraction of the full sky scenario.}
    \label{fig:Cell}
\end{figure}

Fig.~\ref{fig:Cell} shows the SNR, normalized to the full-sky case, given various FoV sizes. The bar plot corresponds to the left axis, showing the SNR contribution from each bin (see Eqn.~\eqref{eqn:lbin}; for ease of presentation, we have normalized all three cases by dividing out the $f_{\rm sky}$ factor). The black solid line shows the full-sky scenario. The right axis and the same black solid line gives the accumulated SNR given various threshold $\ell$. We observe that the SNR drops sharply with limited FoV; in the case of Roman with $\theta_{\rm fov}=0.53~\rm deg$, the fractional SNR is smaller than $10^{-14}$ compared to the SNR of the full-sky case. 

Therefore, only full-sky or nearly full-sky surveys are likely to be sensitive to the angular power spectrum of SGWB. 
If the SNR is sufficiently large, this method can in principle become a probe for modified gravity theories, as is discussed in Refs.~\cite{MihaylovNonC,MihaylovNonE,Qin,Golat}. Just as this method complements PTAs in the frequency domain, it would also cross-check the non-GR effect with the timing residual measurements. To accomplish this goal, it requires a survey with a nearly all-sky FoV, in additional to sufficient cadence and astrometric accuracy.

Lastly, we comment on the detection prospect difference in frequency domain and angular domain. Since $\tilde{h}(f)$ and $C^{E,B}_{\ell}$ are decomposition of the same signal, the total signal power, i.e., $\int d\ln f~h_c^2(f)$, $\sum_{\ell} C^E_{\ell}+C^B_{\ell}$, matches (see Eqn.~\ref{eqn:Cell}). However, the achievable SNRs are limited differently for these two domains; the frequency domain is limited by observational cadence and the angular sensitivity is limited by sky coverage. In particular, the poor angular domain SNR is attributed to the mismatch between the large-scale nature of GW signal and the limited FoV scale.

\section{Conclusion}
\label{sec:conclusion}
In this work, we examine the SGWB detection prospects using astrometric measurements, with discussion of the effects of telescope features on the sensitivity. We consider two especially promising and representative surveys, the Roman GBTD survey and the \textit{Gaia} all-sky astrometric survey, for in-depth discussions and concrete performance forecasts. 

We highlight that whether a uniform deflection signal can be extracted from observational data has a high impact on detector sensitivity. Although absolute astrometry is not required for GW detection, the dedicated telescope design (i.e., two viewing directions) is beneficial for keeping the mean GW signal. For single viewing angle telescopes, it is more likely that we can only use differential deflection data. Thanks to its high-cadence of observations, we found that Roman GBTD survey is sensitive to the SGWB with an SNR of $1.2-120$ for frequency power spectrum analysis, depending on whether the mean signal can be captured, if the background is indeed at the level estimated by current PTA efforts. We found that \textit{Gaia} is able to detect the same signal with an SNR of $2.5$.

We explain the signal loss due to FoV size by binning the power spectrum and disregarding inaccessible bin powers. As the GW signal is intrinsically a large-scale (quadrupole, in particular) signal, and telescopes with the required astrometric accuracy typically have small FoVs, it is unlikely that this detection method is sensitive to the GW angular patterns. To probe the angular power spectrum and potentially observe non-GR signatures, the future candidates should have significant sky coverage. 

As the next generation of photometric telescopes start to go online or go into detailed planning, the possibility of using astrometric measurements to complement existing GW detection strategies is ever more promising. As is shown, Roman and \textit{Gaia} can offer supporting evidence for PTA measurements of the SGWB and probe a complementary SMBHB population in a previously inaccessible frequency range. 

\acknowledgements
We thank the anonymous referees for helpful comments. We thank Andrew Casey-Clyde for helpful discussions. Part of this work was done at Jet Propulsion Laboratory, California Institute of Technology, under a contract with the National Aeronautics and Space Administration. This work was supported by NASA grant 15-WFIRST15-0008 \textit{Cosmology with the High Latitude Survey} Roman Science Investigation Team (SIT).

\textit{Software:} astropy \cite{astropy}, matplotlib \cite{matplotlib}, numpy \cite{numpy}

\bibliographystyle{apsrev4-2}
\bibliography{ref}
\end{document}